# ReVolt: Power Delivery Network-Aware Voltage Droop Control for 2.5D PIM Chiplet Architectures

Vibhanshu Sharma, *Graduate Student Member, IEEE,* Alish Kanani, *Graduate Student Member, IEEE,* Miao Sun, *Graduate Student Member, IEEE*, Janardhan Rao Doppa *Senior Member, IEEE*, Umit Y. Ogras, *Fellow, IEEE,* and Partha Pratim Pande, *Fellow, IEEE*

***Abstract*— Processing-in-memory (PIM)-based 2.5D multi-chiplet platforms are enablers for machine learning (ML) workloads. However, their performance is affected by the power delivery network (PDN), where varying chiplet-level current demand induces spatially and temporally varying voltage droop. These droop events lead to voltage violations, degrades system performance, and impact inference accuracy for ML workloads. In this work, we propose ReVolt, a dynamic operation unit (OU)-based framework for mitigating voltage droop in PIM-based multi-chiplet systems. ReVolt leverages an LSTM-based PDN surrogate to predict per-chiplet supply voltage trajectories at runtime, enabling proactive adjustment of OU size to mitigate droop events. By treating OU size as a control knob, ReVolt regulates chiplet-level current demand while maintaining computational accuracy. This approach prevents voltage droop violations and improves energy-delay product (EDP) while preserving ML model inference accuracy. Experimental results demonstrate that ReVolt prevents voltage droop violations while achieving an average 76 × reduction in EDP compared to existing fixed and dynamic OU-based baselines, without compromising inference accuracy of ML models.**

***Index Terms*—Power delivery network, Voltage droop, Operation unit (OU), Processing-in-memory (PIM), Multi-chiplet systems**

## I. INTRODUCTION

NOVEL 2.5D chiplet platforms provide a new avenue for compact scale-out implementations of emerging compute- and data-intensive workloads. Integrating multiple small chiplets on an interposer offers significantly higher performance and manufacturing yield than conventional planar ICs. Furthermore, multi-chiplet systems achieve better thermal efficiency than 3D ICs and facilitate heterogeneous integration [1]. As a result, large-scale system design on 2.5D platforms has become feasible. However, the performance of a 2.5D system critically depends on the voltage level at each chiplet, which in turn depends on the power delivery network (PDN).

In a 2.5D chiplet system, power originates at off-chip voltage regulators (VRs) and travels through several stages of packaging before reaching each chiplet's supply rails as shown in Fig. 1(a). This multi-stage current path includes the interposer PDN, which behaves as a distributed resistance-inductance-capacitance (RLC) network (Fig. 1(b)). Each segment of the interposer contributes sheet resistance to the lateral current flow, while the package-to-interposer interfaces contribute parasitic inductance to the vertical path [2], [3]. On-die and package-level decoupling capacitances act as local charge reservoirs that suppress high-frequency voltage transients. However, their effectiveness diminishes under sustained current surges that exceed the stored charge [3], [4]. As chiplet count and associated compute density scale up, the aggregate current drawn through this shared network grows proportionally. Hence, the voltage drop accumulated across the PDN directly limits system reliability and performance [4].

The PDN impedance converts fluctuations in chiplet current demand into deviations in chiplet supply voltage, a phenomenon known as *voltage droop*. Two mechanisms contribute to this droop. First, resistive IR-drop, where the sustained DC current drawn by an active chiplet produces a static voltage proportional to the resistance between the VR and that chiplet, pulling its local supply voltage below nominal. Second, is the inductive $di/dt$ drop. Current demand shifts rapidly whenever the computational load changes [2]. The parasitic inductance of the PDN path resists this change, which manifests as a transient voltage undershoot/overshoot. This voltage deviation, which can exceed the steady-state IR-drop, decays only as the inductance allows current to ramp up. This effect is further amplified in multi-chiplet systems. When multiple chiplets transition into compute-heavy phases simultaneously, their current surges are spatially correlated and collectively stress the shared PDN. Consequently, the aggregate droop at each chiplet cannot be predicted from any single chiplet's behavior in isolation [5].

The impact of voltage droop on a multi-chiplet system depends on both the characteristics of the chiplets as well as the application under consideration. For example, in machine learning (ML) workloads such as deep neural networks (DNNs), crossbar-based non-Von-Neumann processing-in-

This work was supported by the US National Science Foundation (NSF) grant CSR-2308530, the Army Research Office under Grant ARO-W911NF-24-1-0240 and the U.S. Department of Energy, Office of Science, Advanced Scientific Computing Research program under project 84245-Democratization of Co-design for Energy-Efficient Heterogeneous Computing (DeCoDe) at Pacific Northwest National Laboratory (PNNL). PNNL is a multi-program national laboratory operated for the U.S. Department of Energy (DOE) by Battelle Memorial Institute under Contract No. DE-AC05-76RL01830. Additional partial support was provided by the Intel CAD SRS Program.
*(Corresponding author: Vibhanshu Sharma).*

Authors' addresses: Vibhanshu Sharma, Miao Sun, Janardhan Rao Doppa, and Partha Pratim Pande, School of Electrical Engineering and Computer Science, Washington State University, Pullman, WA, USA (e-mails: vibhanshu.sharma@wsu.edu; miao.sun@wsu.edu; jana.doppa@wsu.edu; pande@wsu.edu); Alish Kanani, Umit Y. Ogras, Department of Electrical and Computer Engineering, University of Wisconsin-Madison, Madison, WI, USA (e-mails: ahkanani@wisc.edu, uogras@wisc.edu).

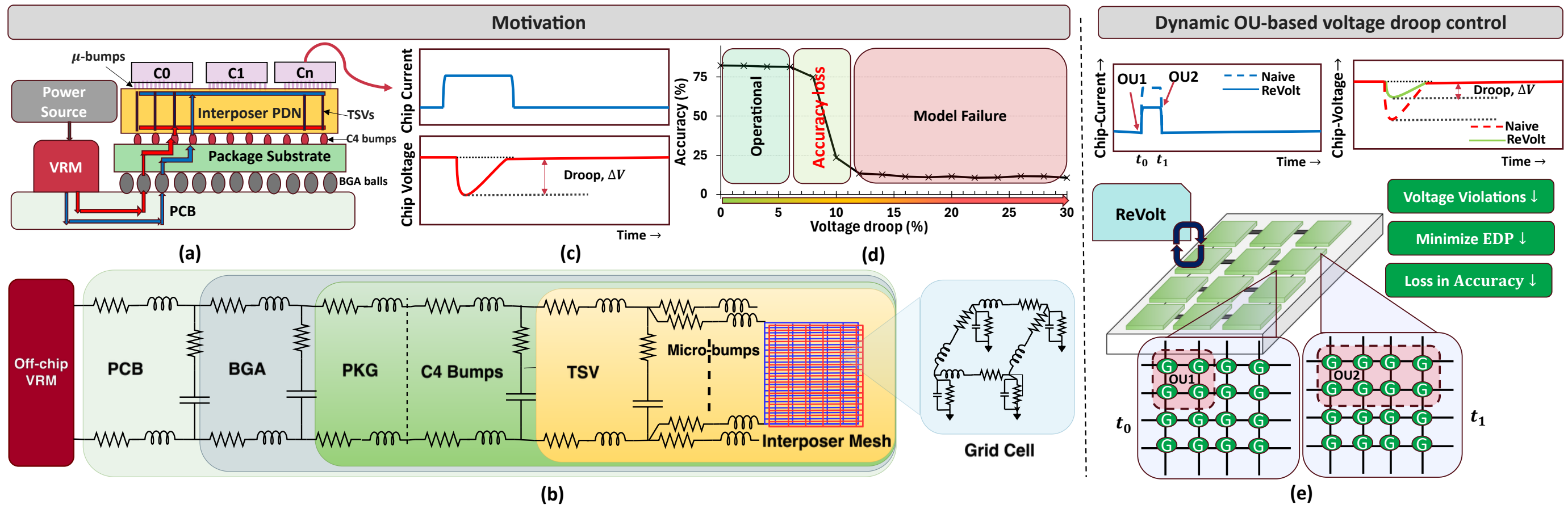

**Fig.1.** **(a)** A conceptual illustration of power-delivery path in 2.5D multi-chiplet architectures. **(b)** Lumped off-chip and distributed on-chip RLC model of the multi-stage PDN. **(c)** Example transient response of the on-chip supply voltage showing a voltage droop ($\Delta V$) induced by sudden workload-driven current surges. **(d)** Showing impact of voltage droop magnitude on ML model inference accuracy. **(e)** PDN-aware runtime OU-based droop control to regulate current demand across the array, minimizing voltage droop while preventing voltage violations, minimizing EDP, and preserving ML model accuracy.

memory (PIM) chiplets outperform traditional Von-Neumann counterparts, such as GPUs [6]. Hence, in this work we consider crossbar-based PIM chiplets as the suitable computing substrate executing ML workloads [7]. However, the crossbar readout makes the computation precision directly dependent on the supply voltage. Voltage droop on the interposer PDN for PIM chiplets therefore affects the ML model correctness, not merely performance headroom [4]. These droop events impose three penalties on the system. First, inference accuracy degrades because analog read circuits and analog-to-digital converters (ADCs) are sensitive to supply voltage, and reduced voltage compresses the output swing, degrading signal resolution at the point of analog-to-digital conversion [4], [8]. Second, reducing the impact of droop imposes a system-level performance cost, as chiplets must either operate at reduced utilization to lower peak current demand, or the nominal supply voltage must be elevated. Both strategies incur energy-delay product (EDP) penalty across all operating conditions [9]. Third, if the supply voltage falls below the minimum operating threshold of the system, device functionality fails, establishing a hard reliability boundary [10]. Addressing these penalties requires a knob that can dynamically regulate PDN current demand. In crossbar-based PIM accelerators, matrix-vector multiplication (MVM) operations are typically executed at a finer granularity than the full crossbar to maintain inference accuracy [11]. This granularity, referred to as an Operation Unit (OU), determines the aggregate current drawn from the PDN. Consequently, the OU naturally serves as a runtime control knob that jointly regulates MVM precision and PDN current demand [11], [12].

We propose a dynamic OU-based control framework, ReVolt, to mitigate voltage droop violations in a PIM-based 2.5D multi-chiplet system. ReVolt employs a long short-term memory (LSTM)-based PDN surrogate to predict per-chiplet supply voltage trajectories, enabling a runtime controller to proactively adjust OU configurations to mitigate droop events. This closed-loop control mechanism effectively regulates PDN current demand, preventing voltage droop violations while minimizing EDP and preserving inference accuracy of ML models. The contributions of this paper are:

1) We propose a runtime OU control framework for mitigating voltage droop violations in a PIM-based 2.5D multi-chiplet system, enabling both proactive and reactive OU adjustments that jointly minimize EDP while preserving inference accuracy of ML workloads.
2) We develop an LSTM-based PDN surrogate model that predicts per-chiplet supply voltage trajectories at runtime, enabling fast droop-aware control.
3) Performance evaluation across multiple DNN and Transformer workloads demonstrates that ReVolt prevents voltage droop violations while achieving on an average $76\times$ lower EDP compared to existing OU baselines, without compromising inference accuracy.

## II. Related Work

### A. PIM-Based Chiplet Architectures and Operation Unit

Several multi-chiplet architectures based on Von-Neumann computing have been proposed for scalable DNN inference [13], [14]. However, DNN inference is dominated by MVM over large weight matrices, and repeated memory access in Von-Neumann systems introduces a performance bottleneck [6]. PIM architectures address this by performing MVM directly within the memory array [6]. Consequently, various architectures have been proposed to scale PIM to large DNN models through chiplet-based disaggregation on silicon interposers [15], [16]. Various network-on-interposer (NoI) topologies have been proposed in the literature to integrate multiple chiplets [17], [18], [19], [20]. To tackle device and crossbar non-idealities in PIM chiplets, MVM operations are computed at a much lower granularity level than a full crossbar, referred to as an OU [11], [12]. A larger OU size increases IR-drop along the signal path in the crossbar array, degrading DNN inference accuracy [12]. To address this, static OU selection at design time has been proposed, where OU is determined based on DNN characteristics such as sparsity etc., [11], [21], [22]. Dynamic OU adaptation has been proposed to recover accuracy under time-varying conductance drift in nonvolatile memory (NVM) cells, minimizing EDP relative to static configurations [12].

### *B. Voltage Droop Mitigation Techniques*

Voltage droop arises from PDN impedance that converts rapid changes in current demand into transient supply voltage undershoots or overshoots [2], [3], [23]. Prior work predominantly addresses voltage droop in the context of conventional Von-Neumann computing systems such as CPUs and GPUs through various mitigation techniques [9], [24]. Guard-banding inflates the nominal supply voltage to absorb worst-case droop, imposing EDP overhead across all operating conditions. Functional-unit throttling is proposed that reduces guard-band overhead by suppressing active compute units during predicted droop windows in GPU architectures [23].

These mitigation techniques rely on various control knobs in a Von-Neumann system. However, in crossbar-based Non-Von-Neumann PIM architectures, one of the suitable control knobs is the OU. The OU configuration directly regulates the aggregate current drawn from the PDN by controlling the number of simultaneously activated word-lines and bit-lines [12]. Prior work has used dynamic OU control to mitigate device-level nonidealities in the crossbar [12]. However, these approaches optimize solely for device-level degradation and do not model PDN coupling or chiplet supply-voltage variation. ReVolt addresses this critical gap by employing an LSTM-based PDN surrogate to predict per-chiplet supply voltage and leveraging dynamic OU-based control to prevent droop violations while minimizing EDP and preserving ML model accuracy.

## III. Problem Formulation

### *A. Voltage Droop Implications in PIM-based Chiplet Systems*

Consider a PIM-based 2.5D chiplet architecture in which DNN layer weights are mapped to PIM crossbars across chiplets. A layer may be mapped to one or more chiplets depending on the layer size and available crossbar resources. The execution granularity of MVM operations on a crossbar is controlled by the OU, parameterized as ($OU_{row}$, $OU_{col}$). Smaller OU configurations activate fewer rows and columns per compute cycle and therefore require multiple cycles to complete the MVM operation of a full crossbar. This increases execution time and consequently EDP. In addition, the power drawn while executing a DNN layer depends on the OU configuration. Further, the chiplet supply voltages are not independent. Instead, the voltage at the $i^{th}$ chiplet is determined by the aggregate current drawn across all chiplets through the interposer PDN impedance and inductive coupling among chiplets. Therefore, a current surge at one chiplet propagates through the interposer PDN (see Fig. 1(b)) and perturbs the supply voltage at other chiplets.

Next, each crossbar column produces an analog current proportional to the MVM result, which the ADC converts to a digital value. As $OU_{row}$ increases, more word-lines are activated simultaneously, which widens the crossbar output current range and requires higher ADC resolution to distinguish output levels [27]. In turn, higher resolution reduces the allowable supply variation per ADC step and tightens the droop margin. Since the ADC reference voltage scales with the supply voltage, voltage droop reduces both the signal range and the separation between ADC quantization levels. Hence, the minimum chiplet supply voltage at which the ADC can correctly resolve all output current levels is defined as the voltage floor, $V_{floor}$. When the supply voltage falls below $V_{floor}$, the ADC conversion error increases, degrading the ML model inference accuracy. This accuracy-driven voltage floor differs from the functional voltage limit below which the circuit fails to operate. Since $V_{floor}$ is generally higher than the circuit's functional limit, operating above it implicitly guarantees both correct circuit operation and no degradation in ML model accuracy. These observations motivate developing a runtime droop-aware OU-based control that regulates current demand while maintaining chiplet supply voltages above the accuracy-driven voltage floor.

### *B. Voltage Droop Control Problem*

Based on the PIM-based 2.5D chiplet architecture discussed above, we now formulate the voltage droop control problem. We consider the system to consist of $K$ chiplets where each chiplet contains $C$ crossbars of size $d \times d$. Let $l$ be the neural layer being executed on chiplet $i$ at time instant $t$. Let $A(t)$ denote the set of these layer-chiplet pairs $(l, i)$. This covers both the cases where a layer can span over multiple chiplets i.e., same $l$ but different $i$ or multiple layers can be on the same chiplet i.e., different $l$, same $i$. For each $(l, i) \in A(t)$, a droop-aware OU configuration, $\theta_{l,i}(t) = (OU_{row}, OU_{col})$ needs to be determined, where both $OU_{row}$ and $OU_{col}$ take integer values in $[1, d]$. The selected configuration must ensure that the aggregate PDN load across all chiplets does not drive any chiplet supply voltage below $V_{floor}$, while minimizing the cumulative EDP. We consider EDP as the relevant performance metric as it captures both energy and execution time in a single parameter. The problem can be formulated as:

$$\min_{\{\theta_{l,i}(t)\}} \sum_t \sum_{(l,i) \in A(t)} EDP\left(\theta_{l,i}(t)\right) \quad (1)$$

$$s.t. \quad V_i(t) \geq V_{floor}\left(\theta_{l,i}(t)\right), \quad \forall (l, i) \in A(t),\ \forall t \quad (2)$$

where $i \in \{1, \dots, K\}$ represents the chiplets in the system and $l \in L$ represents the DNN layers. The voltage floor, $V_{floor}\left(\theta_{l,i}(t)\right)$ and the chiplet supply voltage $V_i(t)$ depends on the OU and interposer PDN impedance respectively. As mentioned above, satisfying constraint in (2) ensures preserving of DNN inference accuracy. The OU configuration selected for one chiplet perturbs the supply voltage of others. In addition, each OU configuration directly affects both EDP and voltage droop behavior. As a result, the search space grows with $|A(t)|$ and the possible $(OU_{row}, OU_{col})$ configurations. These properties make exhaustive search infeasible at runtime.

In summary, the goal is to determine a droop-aware runtime OU control policy that minimizes cumulative EDP while preventing voltage floor violations in a multi-chiplet system where both $OU_{row}$ and $OU_{col}$ influence power requirements and voltage droop behavior.

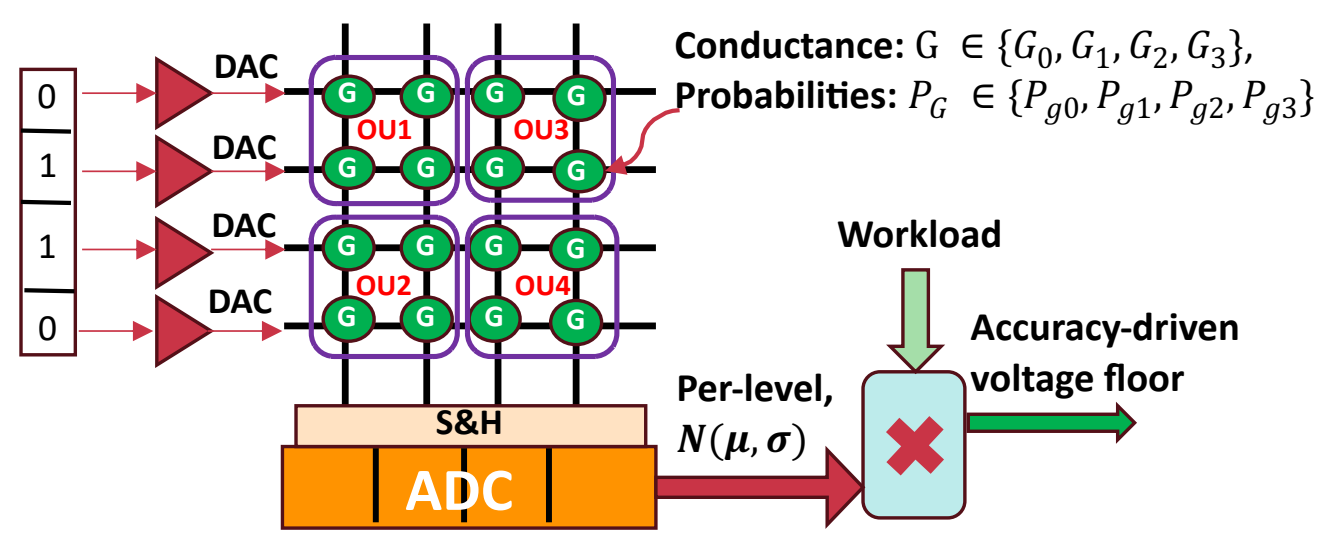

**Fig.2.** Conceptual illustration for probabilistic approach of crossbar current-estimation and the simulation flow to obtain voltage floor.

## IV. Methodology

In this section, we describe each component of the proposed ReVolt framework. First, we present the Current Estimation Model (CEM), which enables ReVolt to estimate the current demand at each chiplet for the next execution window. Next, we describe the LSTM-based PDN surrogate model, which predicts the voltage-droop trajectories for a given chiplet-level current demand. We then introduce the multi-layer perceptron (MLP)-based OU predictor which determines each chiplet's OU configuration based on the predicted voltage droop. Finally, we present the overall ReVolt framework. The frequently used notations are summarized in Table I.

### *A. Current Estimation Model*

To predict voltage droop in a PIM-based multi-chiplet architecture, we first need to estimate the chiplet-level current accurately. For the chiplet system defined in Section III, the current of each chiplet can be obtained by aggregating the current across all crossbars within that chiplet:

$$I_{chiplet} = \sum_C I_{cxb} \tag{3}$$

Each crossbar further includes peripheral circuits such as sample and hold (S&H) unit, ADC, and digital-to-analog converter (DAC), as shown in Fig. 2. Among these, ADCs are a dominant contributor to peripheral current consumption [12]. Hence, the current demand for a crossbar can be estimated as:

$$I_{cxb} \approx I_{arr} + I_{adc} \tag{4}$$

Here, $I_{arr}$ denotes the current drawn by the crossbar array, $I_{adc}$ denotes the ADC current and $I_{cxb}$ denotes the total crossbar current.

We compute the crossbar array current $I_{arr}$ using a probabilistic approach that models the cell conductance as a discrete random variable, whose distribution is determined by the mapped DNN weights. As an example, let us consider a ReRAM-based crossbar with 2 bits per cell storage capacity, yielding four quantized conductance states $G_i = \{G_0, G_1, G_2, G_3\}$, as illustrated in Fig. 2. Let $p_i$ denote the probability that a randomly selected cell is in conductance state $G_i$, computed as:

$$p_i = \frac{N(G_i)}{N_{total}} \qquad where, i \in \{0,1,2,3\} \tag{5}$$

Here, $N(G_i)$ denotes the number of cells in the state $G_i$ and $N_{total}$ denotes the total number of cells. The set $\{p_i\}$ defines a discrete probability distribution over the conductance states $\{G_i\}$. Using this distribution, we compute the mean and variance of the cell conductance:

$$\mu_{cell} = \sum_i p_i G_i \tag{6}$$

$$\sigma^2{}_{cell} = \sum_i p_i (G_i - \mu_{cell})^2 \tag{7}$$

These parameters form the basis for estimating crossbar current under different OU configurations. For a given OU configuration $(OU_{row}, OU_{col})$, the crossbar array current is estimated as:

$$I_{arr} \approx V_{read}.OU_{row}.(OU_{col}.\mu_{cell} + \eta.OU_{col}.\sigma_{cell}) \tag{8}$$

Here, $V_{read}$ denotes the read voltage applied across the word-lines during MVM, and $OU_{row}$ and $OU_{col}$ denote the number of simultaneously activated word-lines and bit-lines respectively. The parameter $\eta \in \{1,2,3,4,5\}$ scales $\sigma_{cell}$ to obtain an upper-bound estimate of $I_{arr}$. In this work, $\eta$ is set to 3, covering approximately 99.7% of the conductance distribution under a Gaussian approximation.

Next, we compute the ADC current $I_{adc}$ to obtain the total crossbar current demand per cycle. The ADC current depends on the column resistance and the ADC bit resolution [28]. The required ADC bit resolution increases with the number of simultaneously activated word-lines and is given by:

$$ADC_{bits} = m + b + \log_2(OU_{row}) - 2 \tag{9}$$

where $m$ denotes the number of input bits processed per cycle and $b$ denotes the number of bits stored per cell. As a result, the ADC current per column increases with $OU_{row}$, while the total ADC current further scales with the number of active columns $OU_{col}$ [28]. Thus, the overall ADC current depends on both OU dimensions and is computed as:

$$I_{adc} \approx \log_2(OU_{row}).OU_{col} \tag{10}$$

This estimation of $I_{cxb}$ enables the ReVolt framework to determine whether the resulting chiplet supply voltage for this current demand will fall below $V_{floor}$ and guides the OU predictor in selecting an appropriate OU configuration.

### *B. PDN Surrogate Model*

Voltage droop exhibits both temporal and spatial dependence due to resistive and inductive effects in the interposer PDN. Consequently, at any instant $t$ the chiplet supply voltage $V(t)$ depends not only on the current $I(t)$, but also on the past chiplet supply voltage $V(t-1)$ and current $I(t-1)$. Further, the supply voltage at a chiplet is also affected by the current demand of other chiplets on the interposer due to the PDN coupling. Therefore, accurate voltage droop prediction must account for both temporal and spatial dependence. To address this, we propose using an LSTM-based recurrent neural model, where the hidden state $h_{t-1}$ and cell state $c_{t-1}$ carry the dependence on past currents and voltages across successive decision instants, as illustrated in Fig. 3 [29]. The LSTM operates as a recurrent block that is applied sequentially at each instant. At every instant, it takes the current input along with the previous hidden and cell states $(h_{t-1}, c_{t-1})$ and produces updated states $(h_t, c_t)$ [29]. This recurrent structure allows the model to propagate information over time, so that the prediction at instant $t$ depends on both the current input and past system behavior, thereby capturing the temporal dependence of voltage droop.

TABLE I
Summary of Frequently Used Notations

| | |
|---|---|
| $\boldsymbol{V_{floor}}$ | Voltage floor |
| $\boldsymbol{(l,i)}$ | Layer-chiplet pair |
| $\boldsymbol{\theta_{l,i}(t)}$ | OU configuration for $(l,i)$ at $t$ |
| $\boldsymbol{N_{Xbar}}$ | Crossbars occupied by layer $l$ on chiplet $i$ |
| $\boldsymbol{I_{cxb}}$ | Total crossbar current |
| $\boldsymbol{I_{chiplet}}$ & $\boldsymbol{V_{chiplet}}$ | Total chiplet current and supply voltage |
| $\boldsymbol{OU_{row}, OU_{col}}$ | Operation unit height and width |
| $\boldsymbol{ADC_{bits}}$ | ADC bit resolution |
| $\boldsymbol{I_{chiplets}(t), V_{chiplets}(t)}$ | Vector representing chiplet current and voltage across all chiplets at instant $t$ |
| $\boldsymbol{\hat{V}_{chiplets}(t)}$ | Vector representing predicted chiplet voltage across all chiplets at instant $t$ |
| $\boldsymbol{E_{comm}}$ & $\boldsymbol{E_{comp}}$ | Communication and computation energy |
| $\boldsymbol{\mathcal{L}at_{comm}}$& $\boldsymbol{\mathcal{L}at_{comp}}$ | Communication and computation latency |
| $\boldsymbol{N_{OU}}$ | Number of OU cycles |

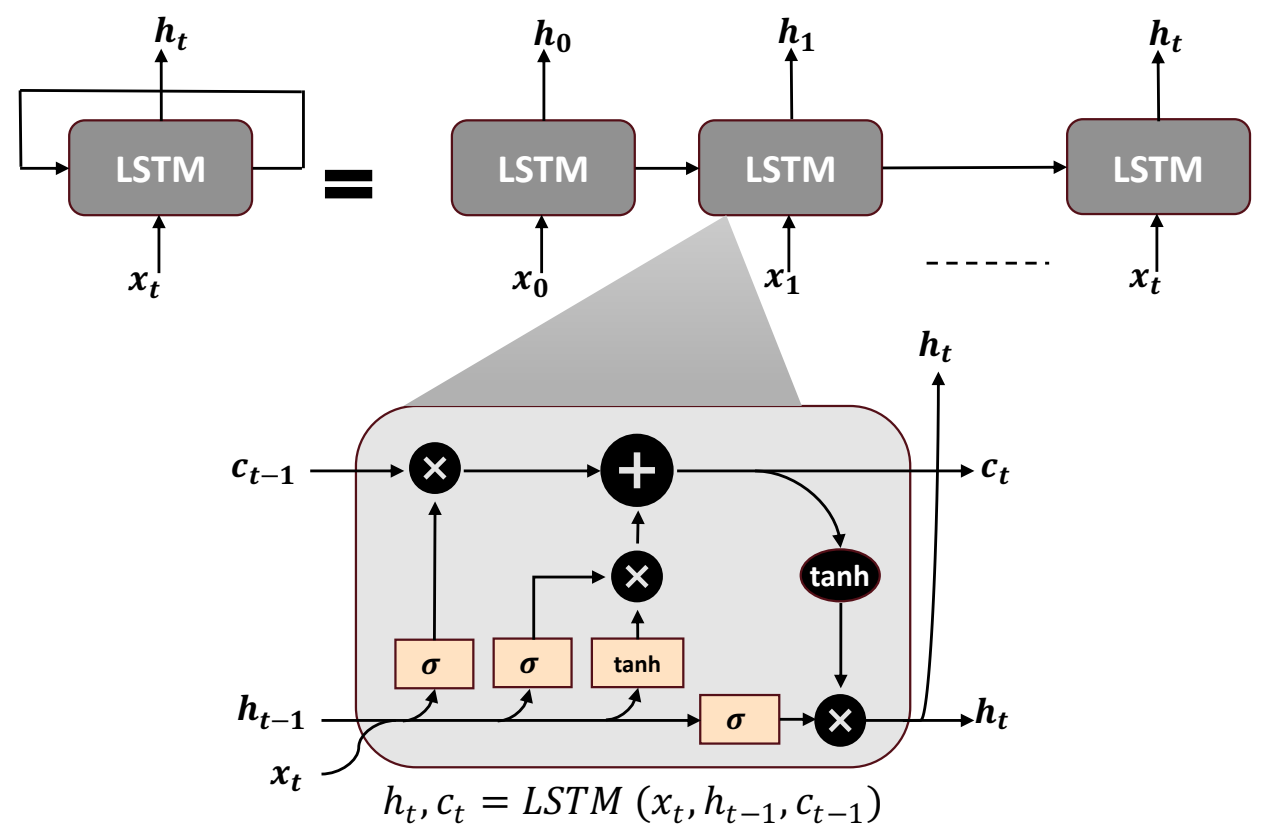

**Fig.3.** LSTM-based model as a recurrent block unrolled across decision instants. The inset shows the internal LSTM cell structure used at each step.

We represent the chiplet current and voltage at an instant $t$ as vectors $I_{chiplets}(t)$ and $V_{chiplets}(t)$, where each vector contains the current and voltage values of all chiplets in the system. To capture spatial dependence, we construct the input $x_t$ by concatenating the chiplet current vector $I_{chiplets}(t)$, and the previous instant chiplet current and voltage vectors, $I_{chiplets}(t-1)$ and $V_{chiplets}(t-1)$, as shown in Eq. 11:

$$x_t = \{I_{chiplets}(t), I_{chiplets}(t-1), V_{chiplets}(t-1)\} \quad (11)$$

Each of these vectors is $K$-dimensional, resulting in an input $x_t \in \mathbb{R}^{3K}$. Since $x_t$ includes current and voltage values from all chiplets simultaneously, the model captures the global spatial coupling across the entire system along with temporal dependencies. The LSTM state update is given by:

$$h_t, c_t = LSTM(x_t, h_{t-1}, c_{t-1}) \quad (12)$$

Using the updated hidden state $h_t$, we predict the chiplet supply voltages as:

$$\hat{V}_{chiplets}(t) = f(h_t) \quad (13)$$

where $\hat{V}_{chiplets}(t)$ denotes the vector of the predicted chiplet supply voltages at instant $t$, with one value per chiplet, i.e., $\hat{V}_{chiplets}(t) \in \mathbb{R}^K$. During runtime, we use this surrogate as a fast decision model for droop prediction in ReVolt.

### C. MLP-Based OU Predictor

We aim to determine per-layer, per-chiplet OU configurations at runtime that minimize EDP while ensuring that each chiplet's supply voltage remains above the voltage floor. The optimal OU configuration depends on workload characteristics (e.g., sparsity and activations of the neural layers) and the resulting voltage droop, which is governed by chiplet current demand and exhibits both temporal and spatial dependence, as captured by the PDN surrogate (Section IV-B). Selecting the OU configuration at runtime is therefore non-trivial. To address this challenge, we employ an MLP-based model that predicts the OU configuration for each layer-chiplet pair $(l, i)$. The model takes as input the workload characteristics, the chiplet current and voltage from the previous instant $(t-1)$, and the chiplet voltage at instant $t$ predicted by the PDN surrogate (Section IV-B). Since the PDN surrogate receives the states of all chiplets simultaneously through its $3K$-dimensional input, the predicted voltage $\hat{V}_{chiplets}(t)$ already encodes the global spatial context across the system.

The predictor operates at the granularity of a single layer-chiplet pair $(l, i)$, as defined in Section III. However, to provide the predictor with supplementary local spatial context, we include features from nearby chiplets. Based on the Green's function formulation for a 2D resistive lattice model of the interposer PDN, the voltage perturbation at a chiplet due to current variations in other chiplets decays with distance [30]. Specifically, the mutual impedance between chiplets decreases with separation, leading to a diminishing influence of far-away chiplets on the local supply voltage [30]. Therefore, we restrict neighbor features used as input to the predictor to the four immediate chiplets, i.e., the north, east, south, and west neighbors with direct connections in the interposer PDN, which captures the dominant local spatial coupling effects around the target chiplet while keeping the predictor input low dimensional.

The input features at instant $t$ for each $(l, i)$ pair consist of the layer sparsity and output activation size, the number of crossbars $N_{Xbar}$ required to map the layer $l$ on chiplet $i$, the chiplet current $I(t)$ obtained from the current estimation model (Section IV-A), and the chiplet supply voltage at the previous instant $V(t-1)$ for chiplet $i$. To capture spatial coupling, we include the mean and maximum voltage drop, and the total current of the neighboring chiplets at instant $(t-1)$, resulting in an input dimension of $\mathbb{R}^8$.

The objective used to generate the training labels is defined as follows. At each decision instant $t$, we evaluate the total EDP using energy $(E)$ and latency $(\mathcal{L}at)$ over all layer-chiplet pairs $(l, i) \in A(t)$, where $A(t)$ denotes the set of layer-chiplet pairs at instant t. Both energy and latency consist of computation $(comp)$ and communication $(comm)$ components:

$$E_{total} = E_{comm} + E_{comp} \quad (14)$$

$$\mathcal{L}at_{total} = \mathcal{L}at_{comm} + \mathcal{L}at_{comp} \quad (15)$$

*Latency:* We evaluate end-to-end latency incurred in ML model inferencing on a crossbar-based PIM architecture. In such architectures, the ADC is the critical component of the pipeline, and its sensing delay scales with bit resolution. The required ADC precision ($ADC_{bits}$) for a crossbar executing the $l^{th}$ neural layer depends on the number of activated word-lines, as discussed in Section IV-A (Eq. 9). Hence, the sensing delay to read each bit-line is proportional to $\log_2(OU_{row})$, where $OU_{row}$ denotes the number of activated word-lines in the OU. Without loss of generality, we consider a PIM chiplet configuration with one ADC per column, as this configuration consumes higher power, representing the worst case scenario for the voltage droop [31] [16]. The OU-based control mechanism in ReVolt can be extended to other PIM configurations by appropriately modifying the energy and latency models and the corresponding voltage-floor constraint, while preserving the overall control framework intact. In this case, the ADC latency for one OU execution will only depend on the $OU_{row}$ as all the OU columns are read in parallel. The computation latency for a layer-chiplet pair $(l, i)$ is obtained by multiplying the latency of one OU with the number of OU cycles $N_{OU}$, where $N_{OU}$ depends on the OU size $(OU_{row}, OU_{col})$ and the layer sparsity [12]. The total computation latency is:

$$\mathcal{L}at_{comp} \approx \sum_{(l,i)\in A(t)} N_{OU}.\,log_2(OU_{row}) \quad (16)$$

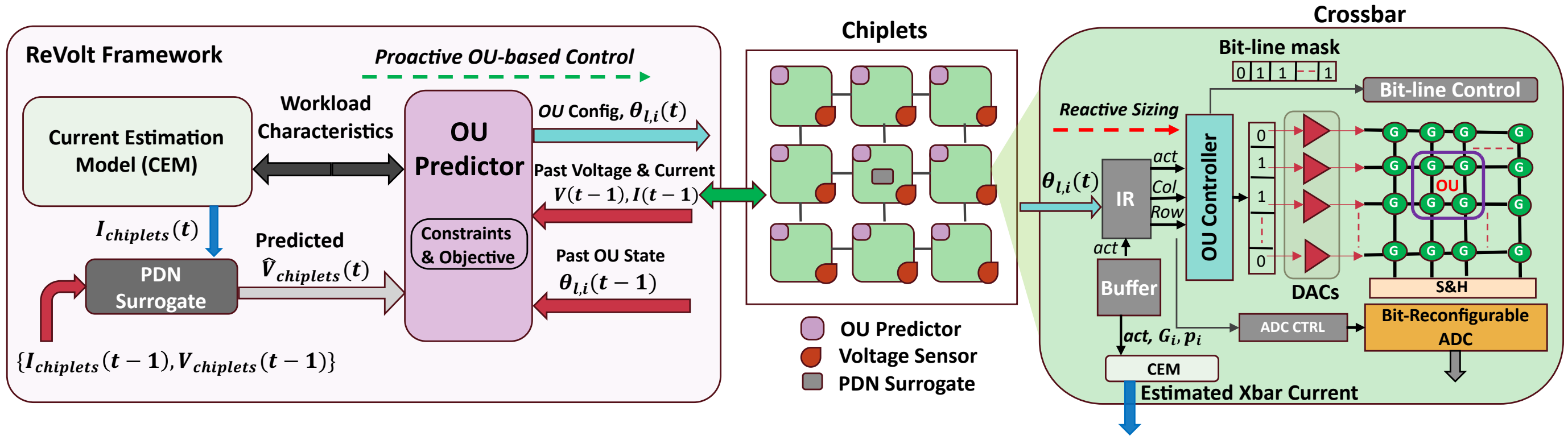

**Fig.4.** Overview of the ReVolt runtime framework. CEM estimates chiplet current demand, which is fed to the PDN surrogate to predict chiplet supply voltages. The OU predictor uses the predicted voltages to select per-layer, per-chiplet OU configurations, which are first latched into an input register (IR) and then applied by the OU controller in the crossbar tile through word-line and bit-line masking.

For a system with $C$ chiplets, let $Q_{ij}$ denote the connectivity between chiplets $i$ and $j$ (1 if a direct communication link exists otherwise 0), $Act_{ij}$ denote the activation volume transferred from chiplet $i$ to $j$, and $h_{ij}$ denote the number of hops between them. The communication latency is:

$$\mathcal{L}at_{comm} \approx \sum_{j=1}^{C}\sum_{i=j}^{C} Q_{ij}.\,Act_{ij}.\,h_{ij} \quad (17)$$

*Energy:* The ADC power consumption scales with bit-precision, as given in Eq. 18 [27]. Since ADC bit-precision ($ADC_{bits}$) is proportional to $OU_{row}$ (Eq. 9), the power consumed in one ADC cycle is proportional to $OU_{row}$, as shown below:

$$P_{ADC} \propto 2^{ADC_{bits}} \propto OU_{row} \quad (18)$$

For a layer-chiplet pair $(l, i)$, the ADC power also depends on the number of crossbars $N_{Xbar}$ required to map the neural layer $l$ on chiplet $i$:

$$P_{ADC}(l, i) \propto N_{Xbar}.\,OU_{row} \quad (19)$$

Using Eq. 16 and Eq. 19, the total computation energy across all layer-chiplet pairs is:

$$E_{comp} \approx \sum_{(l,i)\in A(t)} N_{Xbar}.\,OU_{row}.\,N_{OU}.\,\log_2(OU_{row}) \quad (20)$$

The communication energy for the chiplet-based system is given by:

$$E_{comm} \approx \sum_{j=1}^{C}\sum_{i=j}^{C} Q_{ij}.\,Act_{ij}.\,h_{ij}.\,E_{link} \quad (21)$$

where $E_{link}$ is the energy consumed per bit per hop. The PDN surrogate and the OU predictor together provide the runtime capabilities required by ReVolt. We next describe how these components are integrated into the runtime control loop.

### *D. ReVolt End-to-End Framework*

At each decision instant $t$, ReVolt uses the chiplet current $I(t-1)$ and voltage $V(t-1)$ obtained from on-chip sensors, and the OU assignment $\theta_{l,i}(t-1)$ for all $(l, i) \in A(t)$, as shown in Fig. 4 (left panel). The current estimation model uses the workload at instant $t$, the OU configuration $\theta_{l,i}(t-1)$, and the weight statistics (mean $\mu_{cell}$ and variance $\sigma_{cell}$) calculated from the conductance distribution of cells in the crossbar arrays to compute the chiplet-level current demand $I(t)$ as discussed in Section IV-A.

The PDN surrogate then predicts the chiplet voltages $\hat{V}_{chiplets}(t)$ using $\{I_{chiplets}(t), I_{chiplets}(t-1), V_{chiplets}(t-1)\}$ (see Fig. 4). ReVolt first checks whether the existing OU assignment $\theta_{l,i}(t-1)$ satisfies the OU-dependent voltage-floor constraint as defined in Section III (Eq. 2). To guide this decision, we define voltage headroom as the difference between the predicted per-chiplet voltage $\hat{V}(t)$ and the required voltage floor $V_{floor}$. If the current OU assignment satisfies the voltage-floor constraint, ReVolt computes the headroom and retains the current OU. When sufficient headroom is available, ReVolt proactively evaluates a bounded set of larger OU candidates using the PDN surrogate and selects the feasible candidate that minimizes EDP. If no larger candidate satisfies the voltage-floor constraint, the current OU is retained, consistent with the proactive OU control path shown in Fig. 4 (left panel).

If the current assignment does not satisfy the voltage-floor constraint, or if headroom is insufficient, ReVolt invokes the MLP-based OU predictor to generate a new OU decision $\theta_{l,i}(t)$. The predicted OU is validated using the PDN surrogate against $V_{floor}$. If it violates the constraint, a feasible alternative OU configuration is selected from the candidate set defined in Section III that satisfies the voltage-floor constraint and minimizes EDP. Once all $(l, i)$ pairs are resolved, the final OU assignments are applied for the time instant $t$. The OU controller configures word-line and bit-line activation based on $\theta_{l,i}(t)$ and the ADC applies the corresponding conversion setting by adjusting its bit resolution according to $OU_{row}$, as shown in Fig. 4 (right panel). OU reconfiguration is handled within the crossbar execution pipeline and does not introduce additional pipeline stalls [11]. In case of a voltage-floor violation during runtime, ReVolt switches to a reactive mode. Since the voltage-floor violation has already occurred, the framework prioritizes immediate recovery and bypasses the predictor to avoid additional DNN inference latency. Upon receiving a violation flag from the voltage sensor circuitry, the OU controller searches for a suitable configuration within the discrete OU candidate space. Specifically, the controller employs a binary-search-based recovery mechanism over the supported $OU_{row}$ and $OU_{col}$ configurations, converging to an acceptable OU configuration that satisfies the voltage-floor constraint. The selected OU is then applied through word-line and bit-line masking implemented using combinational logic in the mask update circuitry [11]. This update takes effect within the same execution cycle, enabling an immediate reduction in current demand [11], [12]. Since the reactive control path only downsizes the OU, each corrective action reduces the current demand. Moreover, OU updates are applied at the crossbar granularity by modifying only the active word-lines and bit-lines within an array. Hence, the reactive path progressively mitigates droop without introducing large chiplet-level current transients that could trigger secondary droop events.

ReVolt operates as a closed-loop system that combines the CEM, PDN-surrogate, and OU selection at each decision instant. It uses $I(t-1)$, $V(t-1)$, and $\theta_{l,i}(t-1)$ to select an OU configuration at $t$ that satisfies the $V_{floor}$ constraint (Eq. 2). ReVolt increases the OU size when sufficient headroom is available, invokes the predictor when needed, and enforces safety through validation and selection of a feasible OU. This enables improved EDP while maintaining $V_{floor}$ compliance.

## V. Experimental Results

In this section, we present the experimental evaluation of the ReVolt framework in a multi-chiplet environment. We begin by outlining the experimental setup used for the performance evaluation. Next, we analyze the relationship between voltage droop and DNN inference accuracy under varying voltage-floor constraints. We then evaluate voltage droop in a naïve system (full crossbar activation) to establish the baseline. Further, we evaluate a simple PDN-aware droop-triggered throttling policy to highlight the limitations of purely reactive droop mitigation. We also compare the performance of the MLP-based OU predictor with respect to graph neural network (GNN)- and attention-based models. Moreover, we compare the performance of the LSTM-based PDN surrogate with respect to RLC-circuit model and linear-regression based methodologies. We then evaluate the proposed ReVolt framework and present performance results across a diverse set of workloads and network-on-interposer topologies. We further evaluate the sensitivity of ReVolt to key design parameters. Next, we report the prediction accuracy of the PDN surrogate and analyze the runtime and storage overhead of the framework. Finally, we validate the PDN modeling tool used in this work against an industrial-grade tool, RedHawk to ensure the reliability of our evaluation.

TABLE II
List of Parameters Used for the PDN Model [4]

| Parameter | Value |
|---|---|
| On-die metal resistivity | $1.8 \times 10^{-8}\ \Omega \cdot m$ |
| M1 wire pitch/ width/ thickness | $0.16\ \mu m/\ 0.08\ \mu/\ 0.114\ \mu m$ |
| M2 wire pitch/ width | $0.56\ \mu m/\ 0.28\ \mu m$ |
| M2 wire thickness | $0.506\ \mu m$ |
| M3, M4 wire pitch/ width/ thickness | $39.5\ \mu m/\ 17.5\ \mu m/\ 7\ \mu m$ |
| On-die decoupling capacitance density | $335\ nF/mm^2$ |
| Microbump pitch/ resistance/ inductance | $40\ \mu m/\ 30.9\ m\Omega/\ 11.1\ pH$ |
| C4 bump pitch/ resistance/ inductance | $200\ \mu m/\ 14.3\ m\Omega/\ 11.0\ pH$ |
| Package parallel resistance/ inductance/ capacitance | $0.16\ m\Omega/\ 1.41\ pH/\ 240\ \mu F$ |
| Package line resistance/ inductance | $0.166\ m\Omega/\ 21\ pH$ |

### A. Experimental Setup

**Hardware Architecture details:** Without loss of generality, we consider a PIM-based homogeneous 2.5D chiplet system comprising of 25 identical chiplets arranged in a $5 \times 5$ layout on a passive silicon interposer. Each chiplet occupies $2.5 \times 2.5\ mm^2$ of area, resulting in an active chiplet footprint of $156.25\ mm^2$ [32]. To account for inter-chiplet spacing, routing resources, and package integration, the interposer footprint is approximated as $15 \times 15\ mm^2$, corresponding to a total interposer area of $225\ mm^2$. Each chiplet integrates 300 ReRAM-based $(1T-1R)$ PIM crossbar arrays of size $128 \times 128$, supporting 2-bit/cell weight storage and 1-bit input streaming [31]. Each crossbar includes on-chip buffers to store activations and intermediate results, along with peripheral circuits such as bit-reconfigurable ADCs, DACs, sense amplifiers, and special function units (SFUs) for nonlinear operations [27], [31]. The ADC bit-resolution update signal is applied to the bit-reconfigurable ADC control word through a register at the start of the next conversion cycle and incurs no additional delay [27]. Following existing work, we implement the OU controller needed to activate the appropriate word-lines and bit-lines, consisting of an index decoder, word-line and bit-line mask generator, and multiplexers [12]. The complete OU controller occupies approximately $0.001\ mm^2$ of area and consumes $0.86\ mW$ of power in the 32 nm technology node [11]. Of this, the binary-search logic used in the reactive recovery path contributes to only $111.1\ \mu m^2$ area and $27.4 \mu W$ of power. The resulting controller incurs an area overhead of 4.8% per chiplet [11]. For the ReRAM crossbar, we consider the OU size $(OU_{row}, OU_{col})$ to be constrained by $2^T$, where $T$ is an integer value between 2 and 7 (i.e., 6 discrete values for each dimension). Thus, the binary-search-based reactive path resolves voltage-floor violations in at most 6 feedback iterations ($\lceil \log_2(36) \rceil = 6$). ReVolt operates at an event-driven granularity, where the CEM continuously monitors chiplet current demand and the control framework is invoked only when the predicted demand exceeds a threshold. We undertake a detailed sensitivity analysis of this threshold in Section V-G. A lower threshold increases the number of predictor invocations and can trigger frequent OU transitions, which can introduce abrupt changes in current demand and exacerbate droop events. Conversely, a higher threshold delays control actions and increases voltage-floor violations. As we demonstrate later, the threshold is set at 10%, which balances predictor overhead and control responsiveness while maintaining an effective droop mitigation. The PDN is modeled as a multi-stage RLC electrical network as shown in Fig. 1(b), consisting of chiplet-local on-chip power grids, a shared interposer redistribution network, and the package-board supply path consistent with existing literature [2], [3].

**Tools and Platforms:** We employ CIM-Loop (version 1.0) to model the PIM architecture, including microarchitectural configurations such as row/column drivers, ADCs, memory cells, and digital components including adders and accumulators [33]. CIM-Loop enables accurate data value-dependent modeling for ReRAM-based PIM chiplets, and has been validated against fabricated systems, reporting $\leq 5\%$ difference in throughput and energy-efficiency and $\leq 8\%$ difference in area [33]. Each DNN task is partitioned and mapped onto the multi-chiplet system. The inter-chiplet traffic is generated by the expected activation flow among the neural layers. Using this flow, power traces are constructed considering the mapping and execution order of the neural layers. Additionally, we use BookSim (version 2.0) to model inter-chiplet communication, with the energy per bit, router frequency, and the number of channels configured according to the UCIe protocol [34] [35]. The NoI topology, chiplet count, and inter-chiplet traffic traces are provided as inputs to BookSim, which instantiates routers with virtual channels, credit-based flow control, and per-flit arbitration to estimate communication latency and energy [34].

The voltage traces corresponding to the generated power traces are obtained using the VoltSpot tool (version 2.0) [3]. VoltSpot is a pre-RTL, architecture-level PDN modeling tool that captures transient voltage variations using a fine-grained grid-based electrical model [3]. However, the original version of VoltSpot supports only monolithic 2D and 3D stacked architectures and does not support 2.5D systems. To address this, we modified the tool so that lateral connectivity is restricted within each chiplet. Cross-chiplet interactions are modeled through the physically shared interposer and vertical package connections, including microbump/C4. This preserves chiplet's local IR-drop behavior and enables accurate cross-chiplet voltage coupling. Table II summarizes the physical parameters used for the PDN modeling, consistent with prior works [4]. For validation, VoltSpot is compared against an industrial-grade tool RedHawk [36]. RedHawk is used for power integrity analysis sign-off, including accurate modeling of IR drop and electromigration in IC and chiplet designs [36]. The runtime framework, including the PDN surrogate and OU predictor, is implemented in Python. We validate the CEM by comparing its predicted crossbar current with the exact current computed by summing programmed conductance values within each column for representative DNN weights across all supported OU configurations. The worst-case overestimation is 9%, observed for the smallest OU size ($OU_{row} = 4, OU_{col} = 4$). This conservative estimation introduces an inherent safety margin that absorbs small sensing inaccuracies and biases toward safer OU selections, thereby preserving voltage-floor compliance. We use 8-bit fixed-point precision for the computation and storage of model parameters and activations. All experiments run on an NVIDIA $A40$ GPU.

**Prediction Model Training and Architecture:** To train the PDN surrogate, we generate an ensemble of chiplet-level current-voltage traces using VoltSpot [3]. To improve generalization, the training dataset includes traces collected from multiple workloads and NoI configurations. The resulting dataset enables the surrogate to learn the underlying spatio-temporal characteristics of the PDN. The surrogate is implemented as a single-layer autoregressive LSTM with a hidden dimension of 128. At each decision instant, the model receives a 75-dimensional input vector as discussed in Section IV-B and predicts the voltage trajectory for the next 10-100 timesteps. To train the OU predictor, we generate training labels using the same workload ensemble through an exhaustive offline search over the discrete OU design space. For each sampled state of the system, the framework evaluates all possible OU configurations under the optimization formulation defined in Section III. Candidate configurations that violate the voltage-floor constraint in Eq. 2 are discarded. The remaining feasible configurations are then evaluated using the EDP objective in Eq. 1, where the energy and latency terms are computed using Eqs. 14-21. Among all feasible candidates, the configuration that yields minimum EDP is selected as the optimal OU and used as the training label. The predictor is implemented as a single-hidden-layer MLP with 64 hidden units. The input and output dimensions follow the formulation described in Section IV-C. Through this strategy, the MLP-based OU predictor learns an optimized OU-selection policy, enabling efficient runtime OU selection through lightweight inference. Both the PDN surrogate and OU predictor are trained offline once and amortized across multiple DNN inference runs.

**Workload and Evaluation Metrics:** We evaluate the performance of the proposed framework considering various DNN workloads including ResNet18, ResNet50, MobileNetv3_Large (MNetV3), Inception_v3, and Vision transformer (ViT) and an encoder-only transformer model, BERT-base (BERT) with datasets including CIFAR-100, TinyImageNet, and SWAG (NLP). The transformer architecture is primarily characterized by a self-attention mechanism and a feedforward (FF) network [37]. The attention mechanisms of transformers that require dynamic operand multiplications would necessitate a high frequency of write operations to ReRAM cells [38]. Since ReRAM writes are slow and ReRAMs suffer from limited write endurance, they are not suitable for multi head attention (MHA) computation in transformers, as explained in existing literature [38]. In contrast, the FF network computation is independent of the input sequence length and results in limited number of updates. Hence, we implement the FF network of ViT and BERT model on ReRAM in the proposed architecture.

Additionally, to study the impact of inter-chiplet communication structure on voltage droop, we evaluate multiple existing NoI topologies, including SIAM, HexaMesh, Kite, and Floret for the same system size of 25 chiplets [17], [18], [19], [20]. The SIAM topology serves as the mesh baseline and provides short-range nearest-neighbor connectivity with predictable routing characteristics [19]. HexaMesh is a compact-packing high-fan-out interconnection architecture that improves bisection bandwidth and has inherently larger router ports with star-like connections to its possible neighbors [20]. Similarly, Kite introduces a family of NoI topologies that use diagonal long-range links to decrease the total hop count between chiplets [18]. Floret is a space-filling-curve-based NoI architecture that accelerates CNN workloads by employing a dataflow-aware mapping along the contiguous chiplets [17]. These topologies directly affect the inter-chiplet hop-count, which in turn affects the current demand and voltage droop behavior of the chiplets.

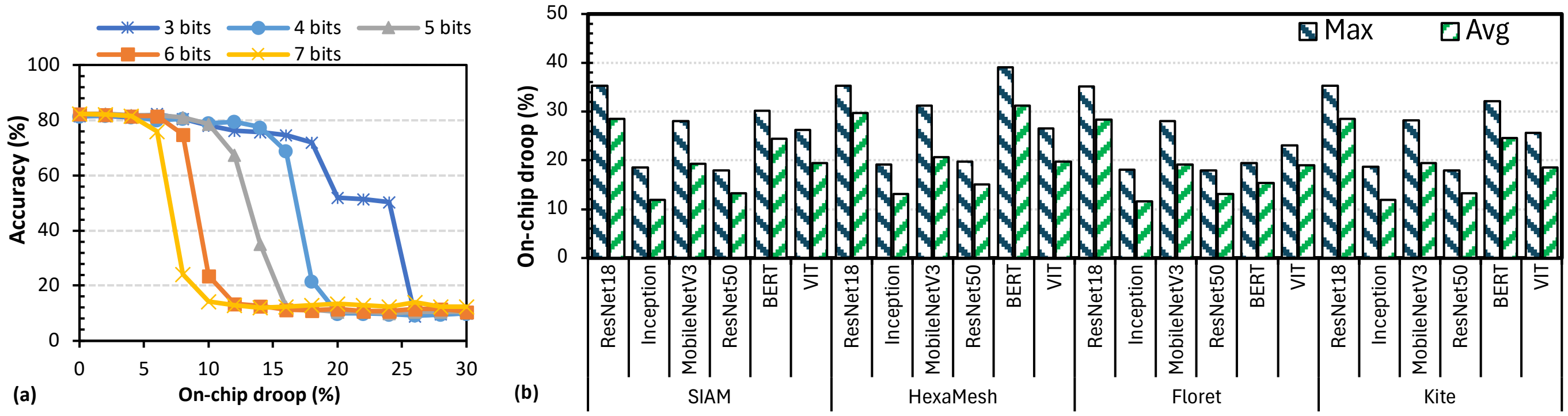

**Fig.5. (a)** Impact of on-chip voltage droop on prediction accuracy for the ResNet18 model with CIFAR-100 under varying ADC precisions (3–7 bits). **(b)** Maximum and average on-chip voltage droop across different ML workloads and NoIs for a naïve system (full crossbar activation).

We evaluate voltage droop as the deviation of the chiplet supply voltage from its nominal value. Droop margin is defined as the difference between the supply voltage and the voltage floor ($V_{floor}$). For droop simulations, $V_{floor}$ captures the impact of droop on predictive accuracy, i.e., voltage constraint violations (Eq. 2) lead to degradation in predictive accuracy, while no violation implies no degradation. These metrics collectively evaluate the effectiveness of the proposed framework in controlling voltage droop, improving system performance, and maintaining DNN prediction accuracy.

### *B. Impact of Voltage Droop on Predictive Accuracy*

ADC resolution determines how well crossbar output levels can be distinguished, which defines the sensing margin of the ADC. Voltage droop reduces this margin by lowering the reference voltage and compressing the separation between quantization levels, thereby introducing conversion errors and degrading inference accuracy, as discussed in Section III. To study this effect, we model the impact of voltage droop on ADC output using a Monte Carlo simulation by injecting droop-induced ADC quantization shifts into the DNN inference pipeline and measuring accuracy across droop levels and ADC precisions. As shown in Fig. 5(a) for the ResNet18 model on the CIFAR-100 dataset, prediction accuracy degrades sharply with increasing voltage droop, with higher-bit ADCs exhibiting faster degradation due to tighter sensing margins. These results show a clear droop threshold beyond which accuracy drops significantly for each ADC configuration. This threshold defines the maximum tolerable droop, and hence the corresponding voltage floor $V_{floor}$, for a given ADC resolution and workload. Maintaining chiplet supply voltage above $V_{floor}$ is therefore necessary to ensure correct inference behavior. Table III lists the computed $V_{floor}$ values for the evaluated workloads and ADC bit resolutions.

### *C. Voltage Droop in Naïve System*

We first evaluate a naïve system where the full crossbar is activated, resulting in higher current demand per cycle (higher power), which in turn leads to increased voltage droop. The analysis is performed using the modified VoltSpot tool discussed in Section V-A. Fig. 5(b) shows the maximum and average on-chip voltage droop across different workloads and NoI topologies considered in this work. These results show that voltage droop depends on both workload characteristics and NoI topology, as they jointly determine the aggregate current demand across chiplets. We observe that among the evaluated NoIs, HexaMesh exhibits the highest droop due to higher inter-chiplet connectivity, resulting in increased router power consumption. At the same time, Floret shows the lowest droop due to fewer inter-chiplet links and reduced power. We further observe that voltage droop reaches up to 39% for the HexaMesh NoI with the BERT-base workload on the SWAG dataset, representing the worst-case experimental scenario. Such high voltage droop leads to voltage-floor violations, degrades predictive accuracy as shown in Fig. 5(a), and renders the system unreliable for correct operation.

### *D. Heuristic Droop-Triggered Throttling*

To isolate the benefit of the learning-based framework, we implement a simple PDN-aware droop-triggered throttling policy. In crossbar-based PIM systems, throttling is naturally implemented by reducing the OU size, which lowers the aggregate current drawn from the PDN. Accordingly, the policy initializes each DNN layer-chiplet pair using the largest OU configuration of (128 × 128). During runtime, whenever the on-chip voltage sensors detect a voltage-floor violation, the controller progressively downgrades the OU according to a precomputed ordering of OU configurations ranked by current-reduction capability. The LSTM-based PDN surrogate is used only to evaluate candidate OUs until the voltage-floor constraint is satisfied. Therefore, the policy remains PDN-aware but does not employ any learned OU-selection mechanism. Fig. 6(a) compares the heuristic policy against ReVolt considering the SIAM NoI and ResNet50 workload as a representative example. The heuristic policy incurs both higher EDP and more voltage-floor violations. Unlike ReVolt, the policy is entirely reactive and does not proactively predict safe OU configurations. Consequently, corrective action is applied only after a droop event, allowing voltage-floor violations to occur before mitigation. Further, as the controller always throttles the OU in response to a violation, execution gradually converges toward smaller OUs, increasing EDP.

TABLE III
DNN Inference Workload Specific $V_{floor}$ Values

| Workload | $V_{floor}$ (%) by ADC resolution | | | | |
|---|---|---|---|---|---|
| | 3-bit | 4-bit | 5-bit | 6-bit | 7-bit |
| ResNet18 | 13.0 | 11.0 | 10.0 | 8.0 | 5.0 |
| Inception | 14.0 | 13.0 | 11.0 | 9.0 | 6.0 |
| MobileNet_v3 | 14.0 | 12.0 | 11.0 | 9.0 | 7.0 |
| ResNet50 | 13.0 | 11.0 | 10.0 | 7.0 | 5.0 |
| BERT | 14.0 | 13.0 | 10.0 | 8.0 | 7.0 |
| ViT | 12.0 | 11.0 | 9.0 | 8.0 | 5.0 |

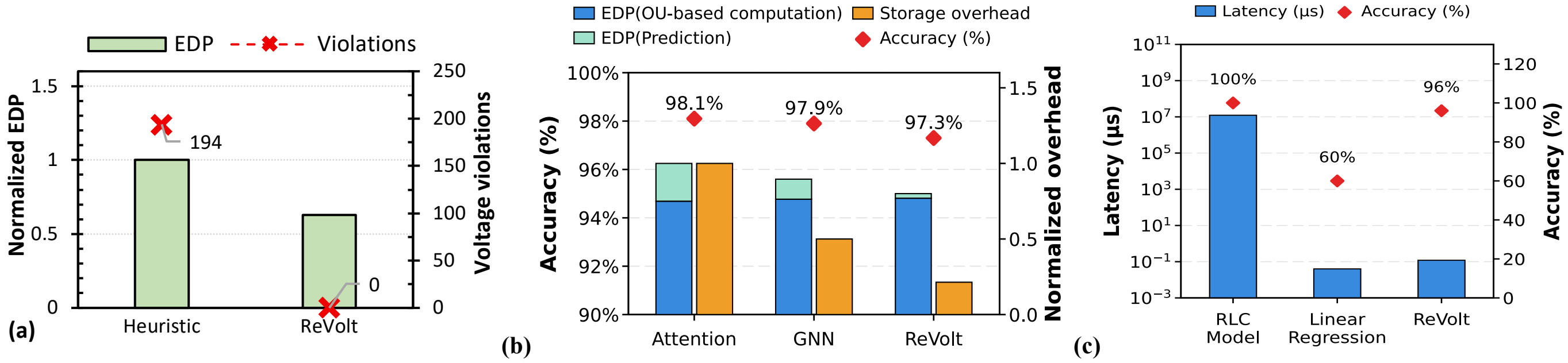

**Fig.6. (a)** Comparison of ReVolt against the heuristic droop-triggered throttling baseline, showing normalized EDP and the total number of voltage violations across the entire simulation for the SIAM NoI and ResNet50 workload. **(b)** Comparison of alternative OU predictor implementations, namely, GNN and Attention based, showing prediction accuracy, normalized EDP breakdown, and storage overhead. **(c)** Comparison of alternative LSTM-based PDN surrogate design approaches, namely, linear regression without recurrent context and RLC-circuit-based model, showing prediction accuracy and latency.

### *E. Predictor Design Tradeoffs*

In this section, we evaluate alternative design strategies for the MLP-based OU predictor and LSTM-based PDN surrogate to quantify the accuracy-overhead tradeoff. We compare the MLP-based OU predictor (Section IV-C) against two spatial-context learning approaches: a GNN-based encoder and an attention-based encoder, each followed by a linear classification head. The GNN-based approach employs a Graph Attention Network (GAT) over the chiplet connectivity, while the attention-based approach uses a transformer-style encoder with a distance-aware attention bias to capture global interactions while preserving spatial locality. Both approaches first construct a contextual representation of the 25-chiplet system state. Specifically, each chiplet $i$ is represented using the feature vector $\{I(t-1),\ V(t-1),\ I(t),\ \hat{V}(t),\ X_{occ}(t)\}$ where $X_{occ}(t)$ denotes the fraction of the crossbars used to map a weight vector on chiplet $i$. The resulting chiplet embeddings are then concatenated with the same local features used by the MLP-based OU predictor (Section IV-C) and the net vector is passed to the classification head to generate the OU decision.

Fig. 6(b) compares the proposed OU predictor against the GAT-based and attention-based alternatives. We report the prediction accuracy, the normalized EDP contributions from OU-based execution and predictor invocation, and the normalized storage overhead. Here, the EDP and storage overhead values are normalized to the attention-based model. We consider ResNet18 running on the SIAM NoI as a representative evaluation scenario. The attention- and GAT-based mechanisms achieve prediction accuracies of 98.1% and 97.9% respectively, compared to 97.3% for the proposed MLP-based OU predictor. However, these limited gains in accuracy come at the cost of substantially higher EDP and storage overheads. As discussed in Section IV-B, the global inter-chiplet coupling (spatial context) is captured by the LSTM-based PDN surrogate and provided to the OU predictor through the predicted voltage $\hat{V}(t)$. The additional spatial context learned by the GAT and attention-based encoders yields only marginal accuracy improvement, while the proposed MLP-based OU predictor achieves comparable accuracy with up to 1.24× lower EDP and 4.67× lower storage overhead.

We next compare the LSTM-based PDN surrogate described in Section IV-B against two alternative approaches: an RLC-circuit-based modeling approach and a linear-regression-based predictor without the recurrent context. For the RLC-based approach, we directly invoke VoltSpot, which models the distributed interposer/package PDN using an RLC network and serves as the ground-truth throughout this work. The linear-regression-based predictor estimates the next voltage state as a linear function of the current voltage and chiplet currents and recursively rolls out the prediction to generate the voltage trajectory. Fig. 6(c) compares the prediction accuracy and runtime overhead of the three approaches. The RLC-circuit-based approach achieves the highest prediction accuracy since it directly evaluates the underlying PDN model. However, each VoltSpot invocation requires 12.3 $s$ on average, making it impractical for runtime OU control. The linear-regression-based predictor incurs minimal computational overhead but achieves only 60% of prediction accuracy, compared to 95.6% for the LSTM-based surrogate. The LSTM-based surrogate provides the best accuracy-overhead tradeoff. By jointly capturing temporal voltage behavior and inter-chiplet coupling, it models the nonlinear PDN dynamics more accurately than linear regression while maintaining significantly lower runtime than RLC-circuit-based modeling approach.

### *F. ReVolt Framework Evaluation*

We first discuss the behavior of the ReVolt framework by analyzing the chiplet supply voltage and OU transitions as shown in Fig. 7. Fig. 7(a) shows the maximum on-chip voltage droop percentage for Chiplet 1 under naïve execution (full crossbar activation), where frequent droop events are observed, highlighted by markers $1-4$. It is evident that the chiplet consistently operates at a lower than the nominal supply voltage. Fig. 7(b) shows the corresponding execution with the ReVolt framework. The same time instances corresponding to markers $1-4$ in Fig. 7(a) are annotated, along with the OU configuration $(OU_{row}, OU_{col})$ selected for each layer-chiplet pair $(l, i)$. At marker 3, where a severe droop event is observed in the naïve case, ReVolt downsizes the OU from (64,64) to (16,32) in anticipation of increased droop, thereby reducing current demand and preventing voltage violation. At markers 1 and 4, where sufficient voltage headroom is available, ReVolt upsizes the OU to improve EDP while satisfying the voltage-floor constraint. Fig. 7(b) shows that ReVolt mitigates these droop events and maintains the supply voltage close to nominal.

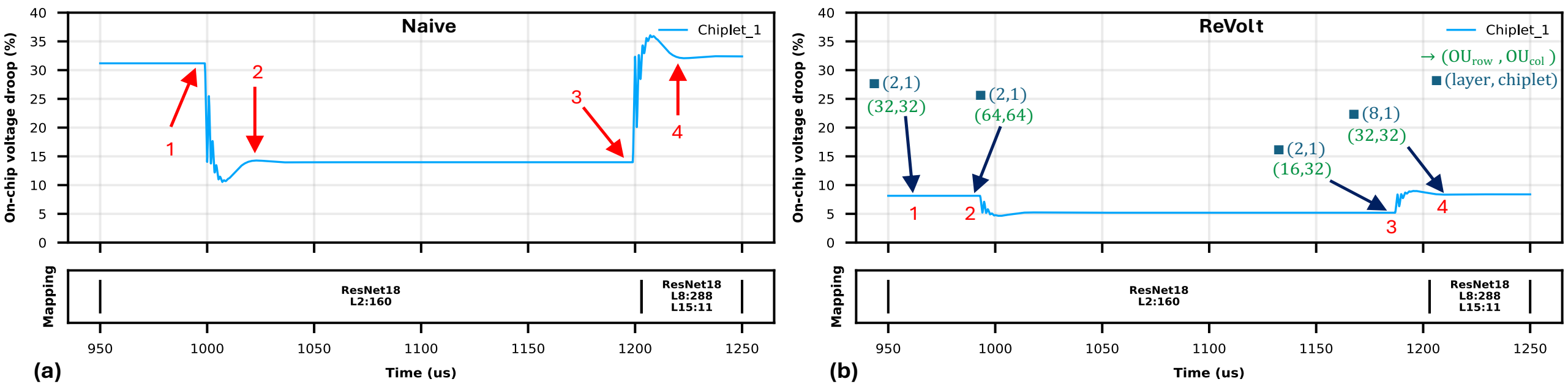

**Fig.7.** On-chip supply voltage of Chiplet 1 and the workload mapping in the format $[layer\text{:}\#crossbars]$ under **(a)** naïve execution (full crossbar activation) and **(b)** the ReVolt framework.

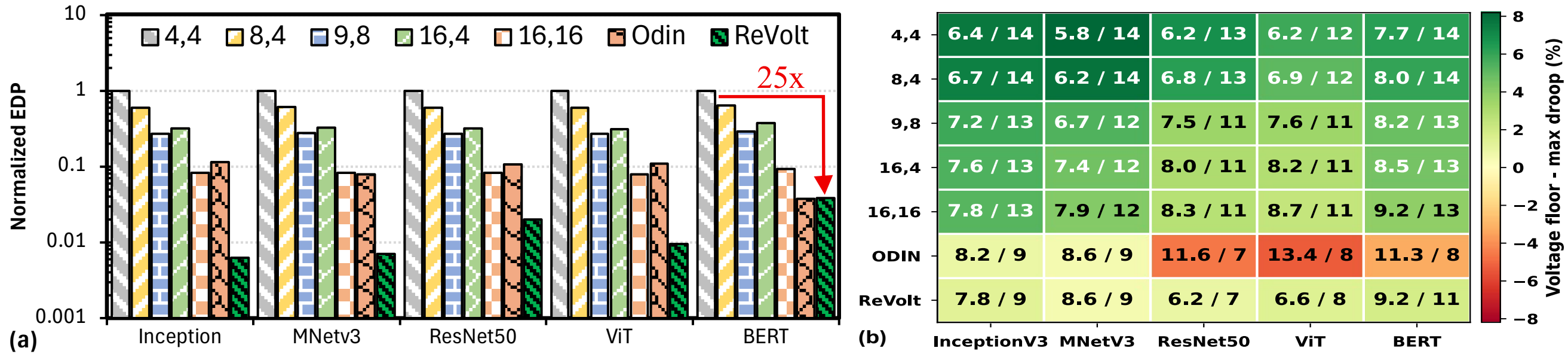

**Fig.8.** **(a)** Normalized EDP comparison (log scale) of ReVolt against fixed and dynamic OU baselines across workloads under the HexaMesh NoI. **(b)** Corresponding voltage droop margin across the same workloads, where each cell reports max on-chip droop / voltage floor.

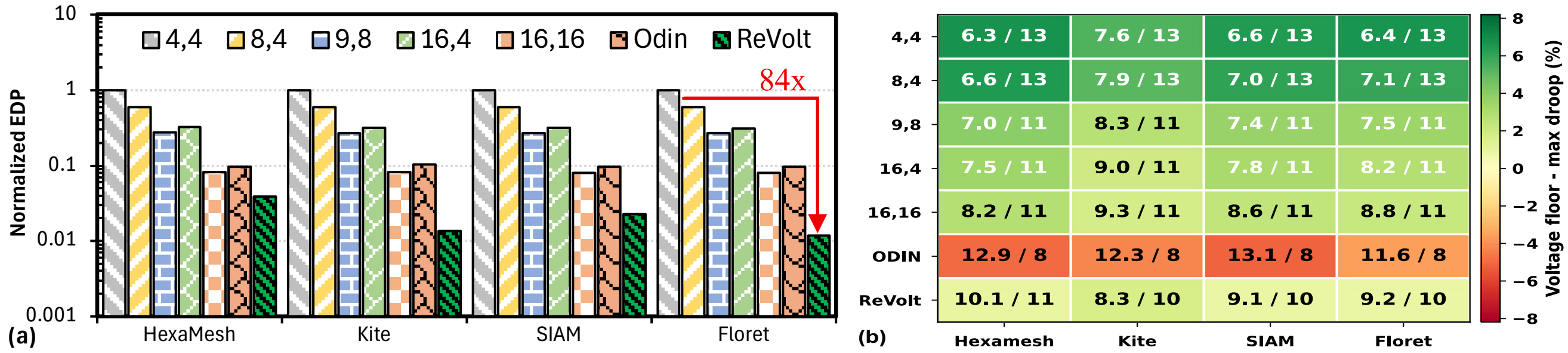

**Fig.9** **(a)** Normalized EDP comparison (log scale) across different NoI topologies for the ResNet18 workload. **(b)** Corresponding droop margin heatmap for the same case. All EDP values are normalized to the fixed 4×4 OU baseline and each cell reports max on-chip droop / voltage floor.

Next, we evaluate ReVolt across diverse ML workloads and different NoI topologies. We consider both fixed and dynamic OU-based baselines. For the fixed OU configurations, we choose the following possibilities: $(OU_{row}, OU_{col}) \in \{(4,4),$ $(8,4), (9,8), (16,4), (16,16)\}$. Each of these configurations is applied uniformly across all $(l, i)$ pairs throughout the whole execution, with no runtime adaptation [11], [21], [22], [39]. We also consider ODIN, where the OU size starts large and progressively transitions to smaller configurations over time to mitigate conductance-drift-induced errors [12]. We model this same dynamic OU-sizing behavior in our evaluation [12].

We first evaluate ReVolt across different ML workloads, considering HexaMesh as the representative NoI. We choose HexaMesh as it shows the highest droop among all the NoI topologies so that it captures the worst-case scenario. Fig. 8(a) shows the normalized EDP in log scale, normalized to the (4,4) baseline. Fig. 8(b) shows the corresponding droop margin, defined as $(V_{floor} - \max_on_chip_droop)$, as a heatmap where color represents its magnitude. A dark green color indicates a higher positive droop margin, while red indicates a negative droop margin, i.e., voltage floor violation. Each cell in the heatmap reports two values in the form $\max_on_chip_droop/V_{floor}$. From Fig. 8(a) and (b), we observe that smaller fixed OU configurations effectively reduce voltage droop, as reflected by large positive droop margins in the heatmap (green-colored cells); however, these configurations overcompensate for droop, indicating that larger OU configurations could have been safely used. Since these baselines are static, they cannot exploit this headroom, resulting in higher EDP due to increased latency at smaller OU sizes, consistent with Section IV. ODIN improves EDP compared to fixed baselines due to dynamic OU sizing. However, its decisions are not PDN-aware, and OU transitions are not aligned with instantaneous droop behavior. This results in voltage violations across workloads, as indicated by negative droop margins. ReVolt, in contrast, is PDN-aware and predicts voltage droop trajectories at runtime. This enables the framework to downsize the OU when droop is predicted and upsize when sufficient headroom exists. Consequently, ReVolt achieves both droop mitigation and performance improvement, with an average EDP reduction of 95 × compared to the (4,4) baseline across workloads.

**Fig.10.** Sensitivity analysis of ReVolt under variations in key design parameters. **(a)** Impact of the predictor-invocation threshold on normalized EDP and voltage-floor violations. **(b)** Impact of voltage-floor constraint selection on normalized EDP and inference accuracy. **(c)** Impact of PDN-surrogate prediction error on normalized EDP and voltage-floor violations.

A key observation is that $OU_{row}$ and $OU_{col}$ affect the system differently. While both influence current and latency, $OU_{row}$ additionally determines ADC resolution and directly impacts the voltage floor. Selecting a smaller $Ou_{row}$ and a larger $OU_{col}$ provides additional voltage margin while reducing latency compared to smaller OU dimensions. ReVolt exploits this tradeoff. This is evident in the BERT workload, where ODIN and ReVolt achieve similar EDP, but only ODIN exhibits voltage violations. ReVolt avoids these violations by selecting a smaller $OU_{row}$ and a larger $OU_{col}$.

We next evaluate ReVolt across different NoI topologies using ResNet18 as a representative workload, shown in Fig. 9(a) and (b). Fig. 9(a) reports normalized EDP in log scale, while Fig. 9(b) shows the corresponding droop margin as a heatmap. Similar trends are observed where fixed small OU baselines overcompensate for droop, resulting in large droop margins but poor performance due to increased latency. ODIN fails to eliminate voltage violations across all NoIs, as indicated by negative droop margins. On the other hand, ReVolt consistently maintains positive droop margins while achieving superior performance. Across NoIs, ReVolt achieves an average EDP reduction of 55 × compared to the (4,4) baseline. The performance gain varies with topology. HexaMesh shows a 25× reduction, while Floret achieves up to 84× reduction. This variation arises from differences in data flow and current distribution across NoIs, which influence droop behavior and available headroom.

### *G. Sensitivity Analysis*

We next evaluate ReVolt under variations in key design parameters, including the current estimation model (CEM) threshold used for predictor invocation, the voltage-floor constraint, and PDN-surrogate prediction error. The results quantify the impact of these parameters on EDP, voltage-floor violations, and overall control effectiveness. We use ResNet18 running on the KITE NoI as a representative example throughout this analysis. As discussed in Section V-A, ReVolt employs an event-driven control policy rather than a fixed periodic update interval i.e., the control framework is invoked only when the predicted demand exceeds a predefined threshold. We first investigate the effect of the predictor-invocation threshold by sweeping its value from 0% to 20%.

Fig. 10(a) reports the normalized EDP, decomposed into two parts, viz., contributions from the OU-based execution and the predictor invocation. It also shows the corresponding voltage-floor violations. At comparatively lower thresholds, the controller invokes the predictor more frequently, reducing the EDP contribution from OU-based execution but increasing that from predictor invocation. For example, at the 0% threshold, the predictor contribution reaches approximately 10%. As the threshold increases, predictor invocations become less frequent, reducing this contribution to approximately 2.37% at the 10% threshold. Although the EDP contribution from OU-based execution increases slightly, the reduction in predictor contribution results in the lowest overall EDP at the 10% threshold. Consequently, ReVolt uses a 10% threshold throughout this work. Further relaxing the threshold reduces the predictor contribution marginally while delaying control decisions. As a result, voltage-floor violations begin to appear beyond the 10% threshold, increasing the EDP contribution from OU-based execution despite the lower predictor contribution.

Next, we evaluate voltage-floor sensitivity by scaling the nominal $V_{floor}$ values by ±10% to represent relaxed and strict operating conditions. Fig. 10(b) reports the resulting normalized EDP. Relaxing the voltage-floor constraint permits the controller to select larger OU configurations, reducing execution latency and lowering EDP. Compared to the strict voltage-floor setting, the relaxed configuration achieves up to 2.3 × lower EDP. However, $V_{floor}$ is determined by ADC sensing requirements and therefore represents a hardware reliability constraint rather than a tunable optimization parameter. Relaxing $V_{floor}$ reduces the sensing margin and can degrade DNN inference accuracy (Fig. 5(a)), whereas the strict setting preserves accuracy but forces the controller toward smaller OU configurations and higher EDP.

Finally, we evaluate the impact of PDN-surrogate prediction error on runtime control effectiveness. To this end, we inject controlled perturbations into the predicted voltage trajectory, varying the error from $-20\%$ to $+20\%$ in 5% increments while keeping the remainder of the runtime control flow unchanged. Fig. 10(c) reports the resulting normalized EDP and voltage-floor violations. Negative perturbations cause the surrogate to underpredict voltage droop, leading the controller to select larger OU configurations (lower latency). Consequently, EDP decreases relative to the nominal case, but voltage-floor violations are observed. In contrast, positive perturbations overpredict voltage droop and bias the controller towards smaller OU configurations. While this eliminates voltage-floor violations, it increases EDP due to more conservative OU selection. At $+20\%$ error, the EDP increases by approximately $1.4\times$ relative to the nominal case. These results indicate that ReVolt responds robustly to both underprediction and overprediction errors, with performance degrading gradually as surrogate error increases.

### *H. Predictor Overhead and VoltSpot Validation*

We first evaluate the storage and energy overhead of the two predictors employed in the ReVolt framework: the PDN surrogate and the OU predictor. The PDN surrogate predicts voltage droop using current and voltage vectors that contain information for all the chiplets, as discussed in Section IV-B. Hence, a single instance of this predictor is sufficient. In our implementation, the LSTM-based PDN surrogate is placed on a chiplet located centrally on the interposer. The required voltage and current inputs are sampled from on-chip sensors and communicated via the UCIe sideband channel, which operates independently of the main data path [35]. It incurs an overall storage overhead of 0.28% considering all the chiplets. For the evaluated 25-chiplet system, the centralized surrogate requires telemetry from the other 24 chiplets, while its own local measurements are already available on-chip. Each remote chiplet contributes three 8-bit telemetry values, namely $I(t-1)$, $I(t)$, and $V(t-1)$, resulting in a small sideband payload. UCIe sideband interfaces operate at 800 MHz (1 bit/cycle), with priority packets enabling deterministic low-latency signaling [35]. As an example, with SIAM as the NoI backbone, telemetry transfer to the centrally placed surrogate requires approximately 28 NoI cycles ($35\,ns$) under the worst-case (corresponding to the maximum source-to-destination hop count in the NoI). This proactive telemetry collection latency is significantly smaller than the control decision interval ( $\sim 1\mu s$) and therefore does not impact the responsiveness of the runtime control loop.

In contrast, the MLP-based OU predictor operates at the granularity of each layer-chiplet pair $(l, i)$, using workload and chiplet features along with neighboring chiplet information, as described in Section IV-C. Therefore, it is instantiated per chiplet, resulting in a system-level storage overhead of 1%. Together, runtime prediction incurs an additional EDP overhead of up to 2.37% over the baseline system. To evaluate prediction accuracy, we perform leave-one-out validation across the workload set [40]. The coefficient of determination $R^2$, which measures the goodness of fit between predicted and ground-truth values, is used for the PDN surrogate. The worst-case $R^2$ is 0.95. For the OU predictor, the prediction accuracy exceeds 97% compared to ground-truth optimal OU configurations. The predictor is trained on an ensemble of workloads with diverse characteristics. The leave-one-out results demonstrate that it generalizes well to unseen DNN workloads. In contrast, the PDN surrogate learns the relationship between chiplet current demand and the resulting voltage behavior for a specific PDN implementation. Consequently, changes to the underlying PDN design or technology node require retraining the surrogate using updated current-voltage traces.

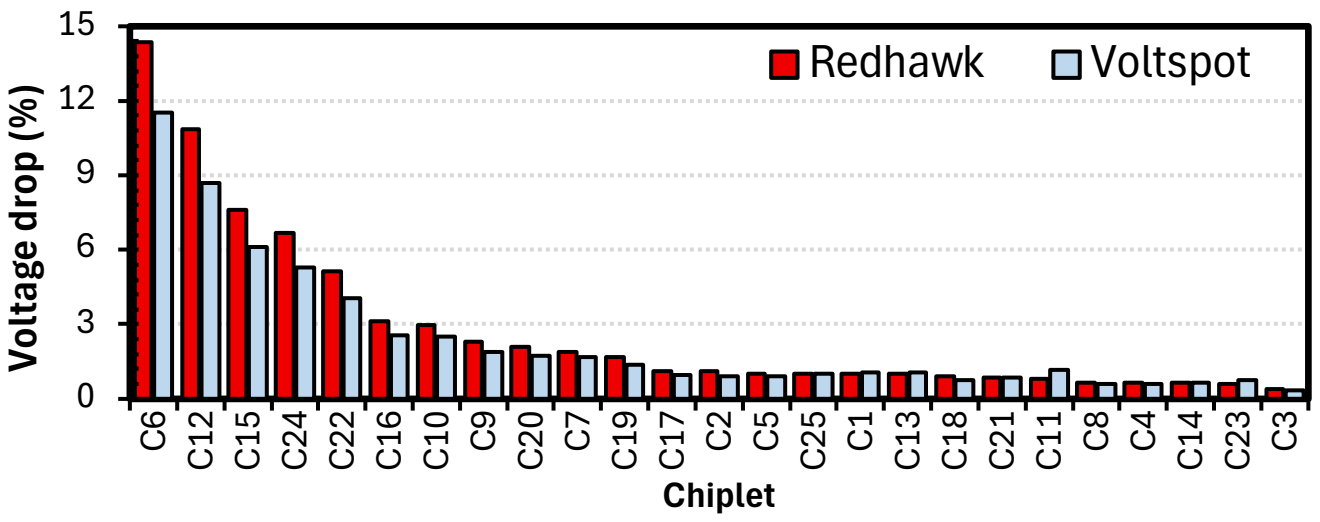

**Fig.11.** Per-chiplet droop comparison between RedHawk-SC and Voltspot.

Next, to validate the accuracy of the VoltSpot tool, we compare its voltage estimates with RedHawk-SC. A reference 2.5D design is implemented in Synopsys ICC2 using GlobalFoundries 22nm CMOS technology, and static IR-drop analysis is performed using RedHawk-SC. The system consists of a homogeneous $5\times 5$ chiplet array integrated on a silicon interposer with four routing metal layers. Since RedHawk does not provide voltage as a function of time, we validate VoltSpot against RedHawk-SC using static voltage drop instead. For a fair comparison, interconnect parameters are matched across setups, chiplet size is scaled from $2.5\times 2.5\ mm^2$ to $0.85\times 0.85\ mm^2$ while preserving layout, and per-chiplet power is assigned with drop measured at power taps. Fig. 11. compares per-chiplet IR drop from RedHawk-SC and VoltSpot, sorted by the drop of each chiplet. The x-axis in the plot shows the chiplet identifier, and the y-axis shows the magnitude of the voltage drop. The voltage drop profiles obtained from RedHawk-SC and VoltSpot show close agreement in magnitude and distribution, with a maximum error of 2.8% and mean absolute error (MAE) of 0.508% at $V_{supply} = 0.8V$, validating accurate drop estimation.

## VI. Conclusion

In PIM-based 2.5D multi-chiplet systems, fluctuations in chiplet current demand across a shared interposer PDN induce voltage droop, which degrades system efficiency and impacts inference accuracy of machine learning models. In this work, we presented ReVolt, a PDN-aware OU-based runtime control framework to mitigate these effects. By leveraging an LSTM-based PDN surrogate to predict voltage droop trajectories and an MLP-based OU predictor, ReVolt proactively adjusts OU size to regulate chiplet current demand while maintaining supply voltages above the accuracy-driven voltage floor. Experimental results across diverse ML workloads and NoI topologies demonstrate that ReVolt prevents voltage droop violations while achieving on an average 76 × lower EDP over existing fixed and dynamic OU-based baselines, without compromising inference accuracy.